\documentclass[fleqn,11pt]{article}

\usepackage[affil-it]{authblk}
\usepackage[english]{babel}
\usepackage{amsmath,amssymb,amsthm,booktabs,tabularx,array}
\usepackage{microtype}
\usepackage[section]{placeins}
\numberwithin{equation}{section}
\numberwithin{table}{section}
\usepackage[hidelinks]{hyperref}

\hypersetup{colorlinks, linkcolor=blue, citecolor=blue, urlcolor=blue}
\usepackage{cite}
\usepackage{mathrsfs}

\usepackage[title,titletoc,toc]{appendix}
\usepackage[margin=1in,left=1in,right=1in]{geometry}
\usepackage{longtable}
\newtheorem{theorem}{Theorem}[section]

\newtheorem{proposition}[theorem]{Proposition}
\newtheorem{corollary}[theorem]{Corollary}
\providecommand{\keywords}[1]{\textbf{\textit{Keywords:}} #1}
\allowdisplaybreaks%to allow subequations to float over to next page
\title{Point Equivalence, Lie Symmetries, and Laurent Compatibility of the Time-Dependent Modified Dispersive Water-Wave System}

\author[1]{Manjit Singh\thanks{Email: manjitcsir@pbi.ac.in, manjitcsir@gmail.com}}
\affil[1]{Yadavindra Department of Sciences, Punjabi University Guru Kashi Campus, Talwandi Sabo--151302, Punjab, India.}

\begin{document}
\maketitle

\begin{abstract}
We classify local point transformations between members of the
\((2+1)\)-dimensional modified dispersive water-wave class with two
nowhere-zero real-analytic time-dependent coefficients. A change of
time reduces the class to a single coefficient \(q\). An invertible
differential substitution identifies this class with a slice of a
previously classified coupled system and supplies a family of point
maps. A direct calculation in the original field variables proves
that the induced coefficient action
\(\widetilde q(F(T))=q(T)/k\), with \(F'>0\) and \(k\ne0\) constant,
exhausts all local point identifications. The order and relative sign
of the first nonzero derivative classify nonconstant pointed
coefficient germs. Constant germs admit arbitrary time generators,
regular nonconstant germs admit one independent time generator
modulo the zero-time-component subalgebra, and critical nonconstant
germs admit none through the base time. On movable analytic
hypersurfaces with \(\phi_x\phi_y\ne0\), the double-pole and
regular-\(V\) Laurent branches both have an obstruction at resonance
\(j=2\) proportional to \(q'\). Both branches are formally compatible
exactly for constant \(q\), which is also the precise criterion for
point equivalence to the standard system.
\end{abstract}

\par\medskip
\par
%\noindent\textbf{PACS numbers:} [codes]

% --------------------------------------------------
% Main text
% --------------------------------------------------
\section{Introduction}
\label{sec:introduction}

The \((2+1)\)-dimensional modified dispersive water-wave (MDWW)
system belongs to a family of coupled nonlinear equations that
includes the Boiti--Leon--Pempinelli (BLP),
Broer--Kaup--Kupershmidt (BKK), and dissipative long-wave systems.
The original BLP formulation and the group analysis of
two-dimensional dispersive long-wave equations provide the
historical setting for these models
\cite{boiti1987integrable,paquin1990group}. Subsequent studies of the
BKK and Broer--Kaup systems have developed symmetry reductions,
conservation laws, and exact solutions, including coherent
structures, soliton interactions, and cusped waves
\cite{ying2001abundant,li2022gramian,kassem2019solitons,zhao2015optimal}.
For the MDWW system itself, previous work has constructed
non-travelling waves, obtained reductions through residual and
direct symmetry methods, and derived rational and interaction
solutions
\cite{li2004families,ma2015symmetry,liang2019consistent,ren2019rational}.
Scaling and hetero-B\"acklund transformations have also been
studied for a constant-coefficient generalization
\cite{gaoGuoShan2023mdww}, and Lie point symmetries and an optimal
system have been reported for a generalized MDWW model
\cite{liu2023liegmdww}.

Allowing the coefficients of a differential system to depend on
time raises the question of which profiles can be related by
changes of variables. For example, differential-invariant methods
have been used to map a variable-coefficient BLP model to a
constant-coefficient form \cite{zhang2021lie}, while other
variable-coefficient BLP and dispersive long-wave studies have
examined exact solutions, symmetry reductions, and conservation
laws \cite{gao2021symbolicblp,jin2023lie}. The point-symmetry
pseudogroup of standard BLP has been determined in
\cite{maltseva2024point}. Of particular relevance to the present
study is the enhanced Lie group classification of three
time-dependent coupled systems in
\cite{kontogiorgisSophocleous2024blp}, where equivalence
transformations are used to reduce the number of arbitrary
coefficient functions before the symmetry classification.

We consider the time-dependent MDWW class studied by Liu and Gao
\cite[Eqs.~(1a)--(1b)]{liu2023painleve}, who trace the model to
Chen's variable-separation study \cite{chen2020soliton}. The
class contains two time-dependent coefficients, \(\alpha(t)\)
and \(\beta(t)\), which multiply the linear spatial terms and
the nonlinear terms, respectively. We assume that both
coefficients are real analytic and nowhere zero on the local
time interval. The change of time \(T=\int^t\alpha(s)\,\mathrm{d}s\)
absorbs \(\alpha\) and leaves the single coefficient
\(q(T)=(\beta/\alpha)(t(T))\). A constant field rescaling maps
every nonzero constant \(q\) to the standard value \(q=2\).
This raises the converse question of whether a general local
point transformation can map a nonconstant \(q\) to the
standard system. It also raises a local classification question
at a critical time of \(q\), where the coefficient itself
cannot serve as a time coordinate.

The normalized MDWW class is closely related to the previously
classified systems in \cite{kontogiorgisSophocleous2024blp}.
The constant-ratio branch can be identified with a coefficient
slice of their Eq.~(6), and an invertible differential
substitution maps the full MDWW class to a slice of their
Eq.~(5). Restricting the transformations in their Theorem~4
to that slice yields a family of point transformations in the
MDWW variables. The inherited maps realize the entire coefficient action that we
ultimately obtain. The direct proof establishes that every MDWW point
map induces that action. In particular, the inherited maps construct
normal representatives at critical times, while the necessity result
shows that their order and relative sign distinguish inequivalent
pointed germs.
The distinction between the equivalence group of a class
and the full collection of transformations between individual
members is developed in
\cite{kingston1998form,popovych2010admissible}. Related applications
to coupled Burgers systems and variable-coefficient
Korteweg--de Vries equations appear in
\cite{kontogiorgis2019enhanced,vaneeva2017groupoid}. The use
of point equivalence to identify variable-coefficient equations
with known integrable models is discussed in
\cite{vaneeva2014equivalence}.

A separate question is how this point-equivalence classification
relates to singularity analysis. The Laurent test for partial
differential equations uses expansions near movable singular
hypersurfaces and requires compatibility at the resonances
\cite{weiss1983painleve,kruskal1997analytic}. Its application
to coupled systems and variable-coefficient models is illustrated
in \cite{garagash1994modification,zhou2022painleve}. For the
time-dependent MDWW system, Liu and Gao obtained the condition
\(\alpha/\beta=\mathrm{const}\)
\cite[Eqs.~(9)--(10)]{liu2023painleve}. We revisit the recurrence
in the normalized variables, treating separately the balance
with a double pole in the second field and the balance in which
that field is regular. The comparison is made on movable analytic
hypersurfaces with \(\phi_x\phi_y\ne0\), where \(\phi\) is a local
defining function. Our conclusions concern formal compatibility
of these Laurent branches.

We show that every admissible local point transformation acts
on the normalized coefficient by
\(\widetilde q(F(T))=q(T)/k\), with \(F'>0\) and \(k\) a nonzero
constant. It follows that local point equivalence to standard
MDWW holds exactly when \(q\) is constant. For a nonconstant
pointed analytic germ, the order of the first nonzero derivative
and its sign relative to the coefficient value at the base time
form a complete pair of invariants. The maximal Lie invariance
algebras distinguish three cases. Constant germs admit an
arbitrary-function family of time generators, regular nonconstant
germs admit one additional time generator, and nonconstant germs
centred at a critical time admit no additional time generator
through that time. On both Laurent branches, the resonance at
index \(2\) forces \(q'=0\), and all resonance conditions are
compatible when \(q\) is constant. Consequently, the formal
Laurent-compatible locus coincides with the standard-equivalence
locus and is strictly contained in the locus admitting additional
time symmetry.

Section~\ref{sec:mdww-class} introduces the normalized class and
its relation to the previously classified systems.
Section~\ref{sec:mdww-point-equivalence} determines the admissible
point transformations and classifies the pointed coefficient
germs. Section~\ref{sec:mdww-lie-classification} gives the maximal
Lie invariance algebras, and Section~\ref{sec:mdww-laurent}
establishes the Laurent compatibility conditions.
Section~\ref{sec:mdww-loci} compares the three loci.
Section~\ref{sec:mdww-exact-solutions} presents exact solutions
on the equivalence and symmetry branches, including pole solutions
on \(y\)-independent hypersurfaces outside the stated Laurent test.
\section{Coefficient normalization and differential equivalence}
\label{sec:mdww-class}

\subsection{The governing system and its normalization}

We consider the $(2+1)$-dimensional modified dispersive water-wave
(MDWW) system in the form studied by Liu and Gao
\cite[Eqs.~(1a)--(1b)]{liu2023painleve},
\begin{equation}
\label{eq:mdww-original}
\begin{aligned}
u_{ty}+\alpha(t)(u_{xxy}-2v_{xx})-\beta(t)(uu_y)_x&=0,\\
v_t-\alpha(t)v_{xx}-\beta(t)(uv)_x&=0,
\end{aligned}
\end{equation}
where \(u\) and \(v\) are functions of \((t,x,y)\). We work locally
and assume that \(\alpha\) and \(\beta\) are real analytic and nowhere
zero on a connected open time interval \(I\). The same two coefficient
functions occur in both equations and define the class under study.

The choice \((\alpha,\beta)=(1,2)\) gives the standard MDWW system,
whose symmetry reductions were studied by Ma, Fei, and Du
\cite{ma2015symmetry}. Gao, Guo, and Shan
\cite[Eqs.~(1a)--(1b)]{gaoGuoShan2023mdww}
studied the subfamily with nonzero constant coefficients and obtained
a scaling transformation and a hetero-B\"acklund construction.

Equivalence transformations provide a way to identify which variable
coefficients can be removed by a change of variables
\cite{vaneeva2014equivalence}. In the present class, a reparametrization
of time normalizes the common coefficient \(\alpha\) to unity.
Fix \(t_0\in I\) and set
\begin{equation}
\label{eq:mdww-time-normalization}
T=T(t)=\int_{t_0}^{t}\alpha(s)\,\mathrm{d}s.
\end{equation}
Since \(T_t=\alpha(t)\ne0\), this change of time has a local inverse
\(t=t(T)\). Define
\[
U(T,x,y)=u(t(T),x,y),\qquad
V(T,x,y)=v(t(T),x,y).
\]
The chain rule gives \(u_{ty}=\alpha(t)U_{Ty}\) and
\(v_t=\alpha(t)V_T\), while the spatial derivatives transform without
additional factors. Substitution into \eqref{eq:mdww-original} and
division of both equations by \(\alpha(t)\) yield
\begin{equation}
\label{eq:mdww-normalized}
\mathcal E_q\colon\qquad
\begin{aligned}
U_{Ty}+U_{xxy}-2V_{xx}-q(T)(UU_y)_x&=0,\\
V_T-V_{xx}-q(T)(UV)_x&=0,
\end{aligned}
\qquad
q(T)=\frac{\beta(t(T))}{\alpha(t(T))}.
\end{equation}
Thus \(\alpha\) determines the new time coordinate, and the ratio
\(\beta/\alpha\), expressed in that coordinate, is the remaining
coefficient. Conversely, every nowhere-zero real-analytic \(q\)
arises locally from \eqref{eq:mdww-original} by taking \(\alpha=1\)
and \(\beta(t)=q(t)\) and identifying \(T\) with \(t\) up to a
translation of the time origin.

For a nonzero constant \(q_0\), the field rescaling
\[
U=\frac{2}{q_0}\widehat U,\qquad
V=\frac{2}{q_0}\widehat V
\]
maps \(\mathcal E_{q_0}\) to the standard system \(\mathcal E_2\)
in the variables \(\widehat U\) and \(\widehat V\). Since
\(q(T(t))=\beta(t)/\alpha(t)\), \(q\) is constant precisely when
\(\alpha/\beta\) is constant. Hence the constant-\(q\) branch
coincides with the branch selected by the Laurent condition of
Liu and Gao \cite[Eqs.~(9)--(10)]{liu2023painleve}.

\subsection{Differential equivalence and induced point transformations}
\label{sec:mdww-prior-classifications}

The class studied by Kontogiorgis and Sophocleous
\cite[Eq.~(6)]{kontogiorgisSophocleous2024blp} contains
\eqref{eq:mdww-original} with \(\beta=2\alpha\) under the coefficient
identification
\[
(a,b,c,d)=(\alpha,2\alpha,\alpha,2\alpha).
\]
Indeed, \((u^2-u_x)_{xy}=2(uu_y)_x-u_{xxy}\). The constant field
rescaling from the preceding subsection maps every member with
\(\beta/\alpha=\mathrm{const}\) to one with \(\beta=2\alpha\).
Thus this identification covers the entire constant-ratio branch
up to point equivalence.

The full time-dependent MDWW class is differentially equivalent to
a subclass of \cite[Eq.~(5)]{kontogiorgisSophocleous2024blp}.
The equivalence is given by the invertible differential substitution
\begin{equation}
\label{eq:mdww-differential-substitution}
w=v-\tfrac12u_y,\qquad v=w+\tfrac12u_y.
\end{equation}
Substituting \(v=w+\tfrac12u_y\) into the first equation of
\eqref{eq:mdww-original} gives
\(u_{ty}=2\alpha w_{xx}+\beta(uu_y)_x\).
Using this identity in \(w_t=v_t-\tfrac12u_{ty}\), together with
the second equation of \eqref{eq:mdww-original}, yields
\begin{equation}
\label{eq:mdww-dlw-form}
\begin{aligned}
u_{ty}-2\alpha(t)w_{xx}-\beta(t)(uu_y)_x&=0,\\
w_t-\beta(t)(uw)_x-\tfrac12\alpha(t)u_{xxy}&=0.
\end{aligned}
\end{equation}
In the notation of \cite[Eq.~(5)]{kontogiorgisSophocleous2024blp},
the coefficient tuple is
\[
(a,b,c,d)=(-2\alpha,-\beta,-\beta,-\tfrac12\alpha).
\]
After the time normalization \eqref{eq:mdww-time-normalization},
this tuple becomes \((-2,-q,-q,-\tfrac12)\). The normalized MDWW
class therefore corresponds to a coefficient slice of the \(b=c\)
subclass treated by Kontogiorgis and Sophocleous
\cite[Theorem~4 and Table~A2]{kontogiorgisSophocleous2024blp}.

A family of the transformations in
\cite[Theorem~4]{kontogiorgisSophocleous2024blp} induces point
transformations of the normalized MDWW class. Let
\(\widetilde T=F(T)\) be a local analytic change of time with \(F'>0\)
and set \(s(T)=\sqrt{F'(T)}\). Taking \(\widetilde y=y\) and setting
both field-scaling constants equal to \(k\ne0\) gives
\begin{equation}
\label{eq:mdww-inherited-point-map}
\begin{aligned}
\widetilde T&=F(T),&
\widetilde x&=s(T)x,&
\widetilde y&=y,\\
\widetilde U&=\frac{k}{s}U-\frac{ks'x}{qs^2},&
\widetilde V&=\frac{k}{s}V,&
\widetilde q(F(T))&=\frac{q(T)}{k},
\end{aligned}
\end{equation}
where primes denote differentiation with respect to \(T\). To see
how the map acts on \(V\), write \(W=V-\tfrac12U_y\). The transformation
in Theorem~4 gives \(\widetilde W=(k/s)W\). Since the additive term
in \(\widetilde U\) is independent of \(y\) and \(\widetilde y=y\),
we have \(\widetilde U_{\widetilde y}=(k/s)U_y\). It follows that
\[
\widetilde V
=\widetilde W+\tfrac12\widetilde U_{\widetilde y}
=\frac{k}{s}\left(W+\tfrac12U_y\right)
=\frac{k}{s}V.
\]
Thus the induced transformation acts on the MDWW coordinates and
fields by a point map.

At any time \(T_*\) with \(q'(T_*)\ne0\), choose \(F=q/k\) and take
the sign of \(k\) to agree with that of \(q'(T_*)\). Then \(F'>0\)
on a sufficiently small neighbourhood of \(T_*\), and
\eqref{eq:mdww-inherited-point-map} gives the local representative
\(\widetilde q(\widetilde T)=\widetilde T\).

In the terminology of \cite{popovych2010admissible}, an admissible
transformation specifies a source equation, a target equation, and
a point map between them. Section~\ref{sec:mdww-point-equivalence}
determines all such transformations for \(\mathcal E_q\), including
maps between germs centred at critical times.
Section~\ref{sec:mdww-lie-classification} compares the resulting
Lie point symmetry generators with the published classification.

\section{Admissible point transformations and coefficient classification}
\label{sec:mdww-point-equivalence}

We determine all admissible point transformations between members
of the normalized class \eqref{eq:mdww-normalized}, using the
terminology of \cite{popovych2010admissible}. The transformations
induced by \cite[Theorem~4]{kontogiorgisSophocleous2024blp} give
the family \eqref{eq:mdww-inherited-point-map}. To establish
completeness, we use the direct method for form-preserving point
transformations \cite{kingston1998form}. We start from a general
point transformation in the MDWW variables and derive its
restrictions by splitting the transformed equations on the
appropriate source jet manifolds.

\begin{theorem}
\label{thm:mdww-admissible-transformations}
Let \(q\) and \(\widetilde q\) be nowhere-zero real-analytic
coefficient germs. Every local point transformation from
\(\mathcal E_q\) to \(\mathcal E_{\widetilde q}\) has the form
\begin{equation}
\label{eq:mdww-general-point-map}
\begin{aligned}
\widetilde T&=F(T),&
\widetilde x&=s(T)x+h(T),&
\widetilde y&=Y(y),\\
\widetilde U&=\frac{k}{s(T)}U
-\frac{k\bigl(s'(T)x+h'(T)\bigr)}{q(T)s(T)^2},&
\widetilde V&=\frac{k}{s(T)Y'(y)}V,
\end{aligned}
\end{equation}
where \(k\ne0\) is constant and
\begin{equation}
\label{eq:mdww-coefficient-action}
F'=s^2,\qquad sY'\ne0,\qquad
\widetilde q(F(T))=\frac{q(T)}{k}.
\end{equation}
Conversely, every transformation satisfying
\eqref{eq:mdww-general-point-map} and
\eqref{eq:mdww-coefficient-action} maps \(\mathcal E_q\) to
\(\mathcal E_{\widetilde q}\).
\end{theorem}

\begin{proof}
For the calculation, write \(t,u,v\) for \(T,U,V\). Consider a
general local point transformation
\[
(\widetilde T,\widetilde x,\widetilde y,\widetilde U,\widetilde V)
=(\tau,X,Y,P,R)(t,x,y,u,v)
\]
on an open set of first jets where its horizontal Jacobian is
invertible.

\paragraph{Jet coordinates and projection.}
Write \(L_1,L_2\) for the source residuals. In \(J^3\), impose
\(L_1=L_2=D_tL_2=D_xL_2=D_yL_2=0\).
The projection onto \(J^2\) is exactly \(L_2=0\). Indeed, over
any second jet with \(L_2=0\), the remaining four relations can
be solved successively for \(u_{xxy}\), \(v_{xxx}\), \(v_{xxy}\),
and \(v_{txx}\). For subsequent splitting it is convenient to
write the same relations as
\begin{align}
u_{ty}&=-u_{xxy}+2v_{xx}+q(uu_y)_x,
\label{eq:mdww-eliminate-uty}\\
v_{tx}&=v_{xxx}+q(uv)_{xx},
\label{eq:mdww-eliminate-vtx}\\
v_{ty}&=v_{xxy}+q(uv)_{xy},
\label{eq:mdww-eliminate-vty}\\
v_{tt}&=v_{txx}+q'(uv)_x+qD_t(uv)_x.
\label{eq:mdww-eliminate-vtt}
\end{align}
In the last relation we substitute the solved expressions for
\(v_t\) and \(v_{tx}\). With this choice of eliminated jets,
\(u_{xxt},u_{xxx},v_{xxy}\) are independent third-order
coordinates. The second-order coordinate \(v_{xx}\) is also free
in this chart, which eliminates \(v_t\) rather than \(v_{xx}\).

\paragraph{Projectability.}
Prolongation of a point map commutes with jet projection. Thus a
map between the third-order systems induces a map between the
projected second-order hypersurfaces. At fixed first jets these
hypersurfaces are affine hyperplanes in the second derivatives.
Their defining affine equations, after pulling the target
equation back to the source, have proportional nonzero linear
parts. In particular, the highest-order part of the pulled-back
target residual comes from \(D_{\widetilde x}^2R\).
If \(D_{\widetilde x}=\sum_i d^iD_i\) and \(z=(u,v)\), its
second-order coefficient tensor, up to an overall sign, is
\(d^id^j C_\mu\), where
\[
C_\mu=R_{z^\mu}-R_{\widetilde T}\tau_{z^\mu}
      -R_{\widetilde x}X_{z^\mu}-R_{\widetilde y}Y_{z^\mu}.
\]
The contact matrix is invertible on the prolongation domain,
so this row is nonzero. The source tensor has only the
\(v_{xx}\) component. Proportionality therefore forces
\(d^t=d^y=0\) and \(C_u=0\). The identities
\(D_{\widetilde x}X=1\) and
\(D_{\widetilde x}\tau=D_{\widetilde x}Y=0\) give
\[
D_{\widetilde x}=\frac{1}{D_xX}D_x,\qquad
D_x\tau=D_xY=0,\qquad
R_u-R_{\widetilde x}X_u=0.
\]
Splitting \(D_x\tau=D_xY=0\) with respect to the free first
derivatives gives \(\tau=\tau(t,y)\) and \(Y=Y(t,y)\).
Multiplying the last identity by \(D_xX\) and splitting with
respect to \(v_x\) yields
\[
R_uX_v-X_uR_v=0,\qquad R_uX_x-X_uR_x=0.
\]
If \((X_u,R_u)\ne(0,0)\), all three columns of
\(\partial(X,R)/\partial(x,u,v)\) are proportional to its
\(u\)-column. Adding the row for \(P\) gives
\[
\operatorname{rank}\frac{\partial(X,P,R)}{\partial(x,u,v)}
\le2.
\]
Since \(\tau\) and \(Y\) depend only on \(t,y\), this contradicts
invertibility of the full point transformation. Hence
\(X_u=R_u=0\).

After substituting for \(v_t\), the coefficient of \(v_{xx}\)
in the transformed second equation is
\[
\bigl(R_v-R_{\widetilde x}X_v\bigr)
\left(D_{\widetilde T}t-\frac{1}{(D_xX)^2}\right).
\]
The first factor is nonzero by the symbol comparison and
invertibility. Since \(D_{\widetilde T}t\) depends only on \(t,y\)
and \(D_xX=X_x+X_vv_x\), splitting with respect to \(v_x\)
gives \(X_v=0\). In the resulting second residual, the
coefficient of \(v_y\) is \(R_vD_{\widetilde T}y\). Thus
\(D_{\widetilde T}y=0\), which gives \(Y=Y(y)\). The
\(v_{xx}\)-coefficient condition now reads \(X_x^2=\tau_t\).

The decisive third-order splits in the first residual are
summarized below. The rows are used successively after the
eliminations \eqref{eq:mdww-eliminate-uty}--\eqref{eq:mdww-eliminate-vtt}.
\[
\begin{array}{c|c|c}
\text{Free jet} & \text{Coefficient in the first residual}
 & \text{Consequence}\\ \hline
u_{xxt}&-P_u\tau_y/(X_x^2\tau_tY')&\tau_y=0\\[1mm]
u_{xxx}&-P_uX_y/(X_x^3Y')&X_y=0\\[1mm]
v_{xxy}&2P_v/(s^2Y')&P_v=0
\end{array}
\]
Here \(P_u\ne0\) by invertibility. The first two coefficients
come from the time and \(x\) components of
\[
D_{\widetilde y}
=\frac{1}{Y'}\left(
D_y-\frac{\tau_y}{\tau_t}D_t
-\frac{X_y-X_t\tau_y/\tau_t}{X_x}D_x\right)
\]
in \(D_{\widetilde x}^2D_{\widetilde y}P\). In the second row
\(\tau_y=0\) has already been imposed. Since \(X_x^2=\tau_t\),
the first two rows give
\(\tau=F(t)\), \(X=s(t)x+h(t)\), and \(F'=s^2\).
At this stage \(D_{\widetilde T}D_{\widetilde y}P\) contributes
\(P_v/(s^2Y')\) to the \(v_{xxy}\) coefficient through
\eqref{eq:mdww-eliminate-vty}. The spatial third derivative
contributes the same amount directly.

\paragraph{Dependent variables and coefficient action.}
Put \(r(t)=\widetilde q(F(t))\). Splitting the second residual
with respect to \(v_x^2\) gives \(R_{vv}=0\), so
\[
R=A(t,x,y)v+B(t,x,y),\qquad A\ne0.
\]
The coefficient of \(u_x\) gives
\[
rsP_u(Av+B)=qAv.
\]
Since \(P_v=0\), splitting this identity with respect to \(v\)
yields
\[
B=0,\qquad P=K(t)u+C(t,x,y),\qquad K=\frac{q}{rs}.
\]
The coefficients of \(uv\), \(v_x\), and \(v\) in the second
residual then give, respectively,
\[
A_x=0,\qquad
C=-\frac{X_t}{rs^2},\qquad
\frac{A_t}{A}=rsC_x=-\frac{s'}{s}.
\]
Using \eqref{eq:mdww-eliminate-uty}, the coefficient of \(v_{xx}\)
in the first residual gives \(A=K/Y'\). Therefore
\((sK)'=0\), so \(sK=k\) for a nonzero constant \(k\).
It follows that \(r=q/k\), which proves the necessity of
\eqref{eq:mdww-general-point-map} and
\eqref{eq:mdww-coefficient-action}.

\paragraph{Verification.}
Let \(L_1,L_2\) denote the source left-hand sides in
\eqref{eq:mdww-normalized}, and let \(\widetilde L_1,\widetilde L_2\)
denote their target counterparts. Substitution of
\eqref{eq:mdww-general-point-map}, with the coefficient action
\eqref{eq:mdww-coefficient-action}, gives the identities
\[
\left.\widetilde L_1\right|_{\eqref{eq:mdww-general-point-map}}
=\frac{k}{s^3Y'}L_1,\qquad
\left.\widetilde L_2\right|_{\eqref{eq:mdww-general-point-map}}
=\frac{k}{s^3Y'}L_2
\]
on unrestricted jets. In both residuals, the terms involving
\(X_t\) and \(s'\) cancel against the contributions of the additive
term in \(\widetilde U\). Since \(sY'k\ne0\), the transformation is
locally invertible.
\end{proof}

To compare the full family with
\eqref{eq:mdww-inherited-point-map}, put \(S=\sqrt{F'}\) and
\(\delta=s/S\in\{-1,1\}\). Apply the inherited map first, denoting
its target coordinates and fields by \((\bar T,\bar x,\bar y,\bar U,\bar V)\)
and its coefficient by \(r(\bar T)=\widetilde q(\bar T)\).
Set \(H=h\circ F^{-1}\). The fixed-coefficient map
\[
(\bar T,\bar x,\bar y,\bar U,\bar V)\longmapsto
\left(\bar T,\delta\bar x+H(\bar T),Y(\bar y),
\delta\bar U-\frac{H'(\bar T)}{r(\bar T)},
\frac{\delta\bar V}{Y'(\bar y)}\right)
\]
is a self-map of \(\mathcal E_r\) by the theorem with identity
time map and \(k=1\). It combines a time-dependent translation,
a \(y\)-reparametrization, and, for \(\delta=-1\), reflection
of \(x\) with the corresponding field signs. Since
\(H'(F(T))=h'(T)/F'(T)\), the composition gives exactly
\eqref{eq:mdww-general-point-map}. Thus the inherited family
realizes every coefficient action in the theorem, and the
necessity proof establishes that the list is exhaustive.

Theorem~\ref{thm:mdww-admissible-transformations} describes the
equivalence groupoid of the normalized class. The same formulas,
with \(q\) treated as an arbitrary element, give generalized
equivalence transformations. For the distinction between
admissible transformations and transformations induced by an
equivalence group, see
\cite{popovych2010admissible,vaneeva2017groupoid}.
The usual equivalence group requires the transformations of the
independent and dependent variables to be independent of \(q\).
In \eqref{eq:mdww-general-point-map}, this requires \(s'=h'=0\),
so \(s\) and \(h\) are constant and \(F\) is affine. The proof
uses the nonvanishing of \(q\) and applies at regular and critical
times, as well as on the constant branch.

\begin{corollary}
\label{cor:mdww-standard-equivalence}
The system \eqref{eq:mdww-original} is locally point-equivalent
to the standard MDWW system if and only if \(\beta/\alpha\)
is locally constant.
\end{corollary}

\begin{proof}
By \eqref{eq:mdww-coefficient-action}, composition with the local
diffeomorphism \(F\) and multiplication by \(1/k\) preserve
constancy of the coefficient germ. Conversely, for a nonzero
constant \(q_0\), take \(F(T)=T\), \(s=1\), \(h=0\), \(Y(y)=y\),
and \(k=q_0/2\). This maps \(\mathcal E_{q_0}\) to \(\mathcal E_2\).
The assertion follows from
\(q(T)=(\beta/\alpha)(t(T))\).
\end{proof}

\begin{corollary}
\label{cor:mdww-coefficient-germs}
Up to local point equivalence with the base time mapped to zero,
the nowhere-zero real-analytic coefficient germs have
representatives
\[
\widetilde q(\widetilde T)=1,\qquad
\widetilde q(\widetilde T)=1+\varepsilon\widetilde T^{\,m},
\quad m\ge1,\quad \varepsilon\in\{-1,1\}.
\]
For a nonconstant germ centred at \(T_0\), \(m\) is the least
positive integer for which \(q^{(m)}(T_0)\ne0\), and
\[
\varepsilon=
\operatorname{sgn}\!\left(\frac{q^{(m)}(T_0)}{q(T_0)}\right).
\]
The pair \((m,\varepsilon)\) is a complete invariant of the
nonconstant pointed coefficient germ.
\end{corollary}

\begin{proof}
Choose \(k=q(T_0)\). For a nonconstant germ, analyticity gives
\[
\frac{q(T)}{q(T_0)}
=1+(T-T_0)^m g(T),\qquad g(T_0)\ne0,
\]
where \(g\) is real analytic. On a sufficiently small
neighbourhood of \(T_0\), \(g\) has constant sign. Set
\[
F(T)=(T-T_0)|g(T)|^{1/m}.
\]
Then \(F\) is real analytic, \(F(T_0)=0\), and
\(F'(T_0)=|g(T_0)|^{1/m}>0\). Taking \(s=\sqrt{F'}\) after
shrinking the neighbourhood if necessary, we obtain
\[
\frac{q(T)}{k}=1+\operatorname{sgn}(g(T_0))F(T)^m.
\]
This gives the stated representative through
\eqref{eq:mdww-coefficient-action}.

For any admissible map taking \(T_0\) to \(\widetilde T_0\),
the coefficient action gives
\[
\frac{\widetilde q(F(T))}{\widetilde q(\widetilde T_0)}
=\frac{q(T)}{q(T_0)}.
\]
Since \(F'(T_0)>0\), composition with \(F\) preserves the order
and sign of the first nonzero term after the constant term.
Thus different pairs \((m,\varepsilon)\) give inequivalent
pointed germs, and germs with the same pair map to the same
representative. For a constant germ, taking \(k=q(T_0)\) and
\(F(T)=T-T_0\) gives \(\widetilde q=1\).
\end{proof}
\section{Classification of maximal Lie invariance algebras}
\label{sec:mdww-lie-classification}

We classify the maximal Lie invariance algebras of
\eqref{eq:mdww-normalized} for nowhere-zero real-analytic coefficient
germs. Earlier symmetry studies include the reductions of the
standard MDWW system by Ma, Fei, and Du \cite{ma2015symmetry}
and the Lie group analysis of a generalized MDWW system by Liu
et al.~\cite{liu2023liegmdww}. For the time-dependent class considered
here, we use Theorem~\ref{thm:mdww-admissible-transformations}.
The local flow of each infinitesimal point symmetry consists of
admissible self-transformations, so differentiation of these maps
gives the general form of the generators. We also verify the
resulting infinitesimal invariance conditions by prolongation.

\subsection{Determining equations and the kernel algebra}

Let
\[
Q=\tau\partial_T+\xi\partial_x+\zeta\partial_y
  +\eta\partial_U+\theta\partial_V
\]
be an infinitesimal point symmetry. For a one-parameter family of
self-transformations in \eqref{eq:mdww-general-point-map}, write
\[
\begin{aligned}
F_\epsilon(T)&=T+\epsilon\tau(T)+O(\epsilon^2),&
h_\epsilon(T)&=\epsilon\lambda(T)+O(\epsilon^2),\\
Y_\epsilon(y)&=y+\epsilon\phi(y)+O(\epsilon^2),&
k_\epsilon&=1+\epsilon\kappa+O(\epsilon^2).
\end{aligned}
\]
On the branch containing the identity, \(s_\epsilon=\sqrt{F_\epsilon'}\),
and hence \(s_\epsilon=1+\tfrac12\epsilon\tau'+O(\epsilon^2)\).
Differentiating \eqref{eq:mdww-general-point-map} at
\(\epsilon=0\) gives
\begin{equation}
\label{eq:mdww-lie-generator}
\begin{aligned}
\tau&=\tau(T),&
\xi&=\tfrac12\tau'x+\lambda(T),&
\zeta&=\phi(y),\\
\eta&=\left(\kappa-\tfrac12\tau'\right)U
      -\frac{\tfrac12\tau''x+\lambda'}{q},&
\theta&=\left(\kappa-\tfrac12\tau'-\phi'\right)V,
\end{aligned}
\end{equation}
where \(\kappa\) is constant. Primes on time-dependent functions
denote differentiation with respect to \(T\), and \(\phi'\)
denotes differentiation with respect to \(y\). Differentiating
the self-map condition \(q(F_\epsilon(T))=q(T)/k_\epsilon\)
gives the classifying equation
\begin{equation}
\label{eq:mdww-lie-classifying}
\tau q'+\kappa q=0.
\end{equation}

Let \(L_1\) and \(L_2\) denote the left-hand sides of the first and
second equations in \eqref{eq:mdww-normalized}, respectively.
For the vector field \eqref{eq:mdww-lie-generator}, direct
prolongation gives the following identities on unrestricted jets:
\begin{equation}
\label{eq:mdww-lie-prolongation}
\begin{aligned}
\operatorname{pr}^{(3)}Q(L_1)
&=\left(\kappa-\tfrac32\tau'-\phi'\right)L_1
  -(\tau q'+\kappa q)(UU_y)_x,\\
\operatorname{pr}^{(2)}Q(L_2)
&=\left(\kappa-\tfrac32\tau'-\phi'\right)L_2
  -(\tau q'+\kappa q)(UV)_x.
\end{aligned}
\end{equation}
Thus \eqref{eq:mdww-lie-classifying} suffices for invariance of
both equations. It is also necessary for a vector field of the
form \eqref{eq:mdww-lie-generator}, since \((UV)_x\) can take
nonzero values on the source equation manifold.

Taking \(\tau=\kappa=0\) gives the subalgebra
\begin{equation}
\label{eq:mdww-lie-baseline}
\begin{aligned}
\mathfrak b_q
&=\bigl\langle X_q(\lambda),Y(\phi)\bigr\rangle,\\
X_q(\lambda)
&=\lambda(T)\partial_x-\frac{\lambda'(T)}{q(T)}\partial_U,\\
Y(\phi)
&=\phi(y)\partial_y-\phi'(y)V\partial_V,
\end{aligned}
\end{equation}
where \(\lambda\) and \(\phi\) are arbitrary. In the fixed
coordinates \((T,x,y,U,V)\), the kernel algebra is the
intersection of the maximal Lie invariance algebras over all
members of the class. It is given by
\begin{equation}
\label{eq:mdww-lie-kernel}
\mathfrak g^\cap
=\bigl\langle\partial_x,Y(\phi)\bigr\rangle.
\end{equation}
Indeed, a common generator must have \(\kappa=0\) by
\eqref{eq:mdww-lie-classifying} on the constant branch. Applying
the same equation to a regular nonconstant coefficient then gives
\(\tau=0\). With \(\tau=\kappa=0\), independence of the
\(\partial_U\)-component from \(q\) forces \(\lambda'=0\).
We call a generator with \(\tau\not\equiv0\) an additional time
symmetry. The subalgebra \(\mathfrak b_q\) contains all generators
with zero time component.

\subsection{Classification at regular and critical times}

\begin{theorem}
\label{thm:mdww-maximal-lie-algebras}
Let \(q\) be a nowhere-zero real-analytic germ centred at \(T_0\).
The maximal Lie invariance algebra \(\mathfrak g_q\) of
\(\mathcal E_q\) is given by the following cases.
\begin{enumerate}
\item If \(q=q_0\) is constant, then
\[
\mathfrak g_q
=\bigl\langle\mathfrak b_q,D_{q_0}(\tau)\bigr\rangle,
\]
where \(\tau(T)\) is arbitrary and
\begin{equation}
\label{eq:mdww-lie-constant}
D_{q_0}(\tau)=
\tau\partial_T+\tfrac12\tau'x\partial_x
-\left(\tfrac12\tau'U+\frac{\tau''x}{2q_0}\right)\partial_U
-\tfrac12\tau'V\partial_V.
\end{equation}

\item If \(q'(T_0)\ne0\), then
\[
\mathfrak g_q=\bigl\langle\mathfrak b_q,Z_q\bigr\rangle,
\]
where, with \(\rho=q/q'\),
\begin{equation}
\label{eq:mdww-lie-regular}
Z_q=
\rho\partial_T+\tfrac12\rho'x\partial_x
-\left[\left(1+\tfrac12\rho'\right)U
+\frac{\rho''x}{2q}\right]\partial_U
-\left(1+\tfrac12\rho'\right)V\partial_V.
\end{equation}
For the pointed representative \(q=1+\varepsilon T\), with
\(\varepsilon\in\{-1,1\}\), this generator is
\begin{equation}
\label{eq:mdww-lie-regular-canonical}
Z_{1+\varepsilon T}
=(T+\varepsilon)\partial_T+\tfrac12x\partial_x
-\tfrac32(U\partial_U+V\partial_V).
\end{equation}

\item If \(q\) is nonconstant and \(q'(T_0)=0\), then
\[
\mathfrak g_q=\mathfrak b_q.
\]
The pointed representatives are \(q=1+\varepsilon T^m\), with
\(m\ge2\) and \(\varepsilon\in\{-1,1\}\).
\end{enumerate}
\end{theorem}

\begin{proof}
If \(q=q_0\ne0\), equation \eqref{eq:mdww-lie-classifying} gives
\(\kappa=0\) and leaves \(\tau\) arbitrary. Substitution into
\eqref{eq:mdww-lie-generator} gives
\eqref{eq:mdww-lie-constant} together with
\eqref{eq:mdww-lie-baseline}.

If \(q'(T_0)\ne0\), shrink the time neighbourhood so that
\(q'\) is nowhere zero. Then \(\tau=-\kappa q/q'\).
Taking \(\kappa=-1\) and setting \(\lambda=\phi=0\) gives
\eqref{eq:mdww-lie-regular}. Every generator is therefore a
constant multiple of \(Z_q\) plus an element of
\(\mathfrak b_q\). For \(q=1+\varepsilon T\), one has
\(\rho=T+\varepsilon\), giving
\eqref{eq:mdww-lie-regular-canonical}.

Suppose that \(q\) is nonconstant and \(q'(T_0)=0\).
Evaluating \eqref{eq:mdww-lie-classifying} at \(T_0\) gives
\(\kappa q(T_0)=0\), so \(\kappa=0\). Analyticity and
nonconstancy of \(q\) imply that the zeros of \(q'\) are
isolated. Thus \(\tau q'=0\) gives \(\tau=0\) away from
those zeros and, by continuity, on the whole germ. Hence
\(\mathfrak g_q=\mathfrak b_q\). The pointed representatives
in the two nonconstant cases follow from
Corollary~\ref{cor:mdww-coefficient-germs}.
\end{proof}

At \(q_0=2\), the families \(D_2(\tau)\), \(X_2(\lambda)\),
and \(Y(\phi)\) generate time reparametrizations, time-dependent
\(x\)-translations, and \(y\)-reparametrizations of the standard
MDWW system, respectively.

To compare our generators with
\cite[Table~A2]{kontogiorgisSophocleous2024blp}, set
\(W=V-\tfrac12U_y\). Since \(D_y\tau=D_y\xi=0\), the first
prolongation of \eqref{eq:mdww-lie-generator} gives
\[
\operatorname{pr}^{(1)}Q(U_y)
=D_y\eta-U_yD_y\zeta
=\left(\kappa-\tfrac12\tau'-\phi'\right)U_y.
\]
Consequently,
\[
\operatorname{pr}^{(1)}Q(W)
=\left(\kappa-\tfrac12\tau'-\phi'\right)W.
\]
The induced generators are therefore point vector fields in
the variables of the published class.

For comparison with the \(b=c\) classification in
\cite[Table~A2]{kontogiorgisSophocleous2024blp}, the MDWW slice has
\(a/c=2/q\) and \(d/c=1/(2q)\). For \(q=1+T\), the regular
generator in the published field variables \((U,W)\) is
\[
(T+1)\partial_T+\tfrac12x\partial_x
-\tfrac32(U\partial_U+W\partial_W).
\]
A necessary determining equation for a candidate on this slice
is \(2\xi_x=\tau'\), by \eqref{eq:mdww-lie-generator}.
In particular, the combination
\(\tau=(a/c)/(a/c)'=-(1+T)\), \(\xi=x/2\) gives
\(2\xi_x-\tau'=2\), independently of the field components.
This provides a direct check on any specialization of a
candidate generator to the slice.

\subsection{Time symmetry and equivalence to standard MDWW}

By Corollary~\ref{cor:mdww-standard-equivalence}, the coefficient
germs that are point-equivalent to standard MDWW are precisely
the constant ones. Theorem~\ref{thm:mdww-maximal-lie-algebras}
shows that an additional time symmetry exists for both constant
germs and regular nonconstant germs. Thus the former set is a
proper subset of the latter. For example, \(q=1+T\) near
\(T=0\) admits the time generator
\[
(T+1)\partial_T+\tfrac12x\partial_x
-\tfrac32(U\partial_U+V\partial_V),
\]
while its coefficient germ is nonconstant and therefore
point-inequivalent to standard MDWW.

For a nonconstant coefficient germ centred at a critical time,
the additional time generator exists locally at nearby regular
times. Any extension of this generator through the critical time
would satisfy \eqref{eq:mdww-lie-classifying} with the same nonzero
\(\kappa\) as on the regular neighbourhood. Evaluation at the
critical time forces \(\kappa=0\), which rules out such an extension.
Equivalently, the quotient \(\mathfrak g_q/\mathfrak b_q\) is
one-dimensional on a regular nonconstant germ and zero on a
nonconstant germ centred at a critical time.
\section{Formal Laurent analysis}
\label{sec:mdww-laurent}

We apply the Weiss--Tabor--Carnevale singularity analysis for PDEs
\cite{weiss1983painleve} to the normalized system
\eqref{eq:mdww-normalized}. We consider integer-pole expansions
near movable analytic hypersurfaces \(\phi(T,x,y)=0\) with
\(\phi_x\phi_y\ne0\). The calculation is carried out on a local
complexification where \(q\) is holomorphic and nowhere zero.
We examine both the branch in which \(V\) has a double pole and
the branch in which \(V\) is regular.

\subsection{Leading balances and resonance matrix}

By the implicit function theorem, the singular hypersurface can
be written locally in Kruskal form as
\[
z=x+\psi(T,y)=0,\qquad P=\psi_y\ne0,\qquad H=\psi_T.
\]
We seek formal Laurent expansions
\begin{equation}
\label{eq:mdww-laurent-series}
U=\sum_{j=0}^{\infty}a_j(T,y)z^{j-1},\qquad
V=\sum_{j=0}^{\infty}b_j(T,y)z^{j-2}.
\end{equation}
The expansion for \(V\) allows a regular branch when its
negative-power coefficients vanish.

To see that these expansions cover the possible integer-pole
orders on the stated hypersurfaces, first suppose that \(U\) is
regular and \(V\) has a pole. Then \(V_{xx}\) is the unique most
singular term in the second equation. If \(U\) has a pole of
order greater than one and \(V\) has a pole, \((UV)_x\) is the
unique most singular term in that equation. If \(U\) has a pole
of order greater than one and \(V\) is regular, \((UU_y)_x\)
is the unique most singular term in the first equation.
Thus \(U\) must have a simple pole. If \(V\) also has a pole,
balancing \(V_{xx}\) with \(q(UV)_x\) requires \(qa_0=n+1\),
where \(n\ge1\) is the pole order of \(V\). For \(n>2\),
\(V_{xx}\) is the unique most singular term in the first
equation. For \(n=1\), the order-\(z^{-4}\) balance in that
equation instead gives \(qa_0=-2\), which is incompatible with
\(qa_0=n+1\). Hence any pole in \(V\) has order two.

At order \(z^{-4}\), the two equations give
\begin{equation}
\label{eq:mdww-laurent-leading}
-6Pa_0-12b_0-3qPa_0^2=0,\qquad
3(qa_0-2)b_0=0.
\end{equation}
Since \(a_0\ne0\), the two leading balances are
\begin{equation}
\label{eq:mdww-laurent-branches}
\begin{array}{c|cc}
\text{Branch} & a_0 & b_0 \\ \hline
\text{Double pole in } V & 2/q & -2P/q\\[1mm]
\text{Regular } V & -2/q & 0 .
\end{array}
\end{equation}
Liu and Gao \cite[Eqs.~(4)--(7)]{liu2023painleve} assumed
nonzero leading coefficients in both fields. In the Kruskal
coordinate \(\phi=z\), their leading coefficients become
\(a_0=2/q\) and \(b_0=-2P/q\), the first balance in
\eqref{eq:mdww-laurent-branches}. Their resonance indices
also agree with those below for that balance. The regular-\(V\)
balance has \(b_0=0\) and \(a_0=-2/q\), so it falls outside
their nonzero-leading-coefficient ansatz.
On the compatible regular branch, the recurrence below gives
\(b_1=b_2=0\), so \(V=b_3z+O(z^2)\).

Let \(L_1,L_2\) denote the left-hand sides of
\eqref{eq:mdww-normalized}. At index \(j\), we set
\([z^{j-4}]L_i=0\) for \(i=1,2\), using
\[
D_x=\partial_z,\qquad
D_y=\partial_y+P\partial_z,\qquad
D_T=\partial_T+H\partial_z.
\]
Here \(\partial_y\) and \(\partial_T\) differentiate the
coefficient functions at fixed \(z\). In particular, the
recurrence retains the time derivatives of the leading
coefficients \(a_0\) and \(b_0\).

The matrix multiplying \((a_j,b_j)^{\mathsf T}\) is
\begin{equation}
\label{eq:mdww-laurent-matrix}
M_j=
\begin{pmatrix}
P(j-3)(j-2)(j-1-qa_0) & -2(j-3)(j-2)\\
-qb_0(j-3) & -(j-3)(j-2+qa_0)
\end{pmatrix}.
\end{equation}
For either balance in \eqref{eq:mdww-laurent-branches},
\[
\det M_j=-P(j+1)(j-2)(j-3)^2(j-4).
\]
The resonances are therefore \(-1,2,3,3,4\). The resonance
\(-1\) corresponds to the arbitrary singular manifold.
At the double resonance \(j=3\), \(M_3\) is the zero matrix,
so compatibility requires both residual coefficients to vanish.

\subsection{Compatibility at the resonances}

The nonsingular recurrence at \(j=1\) gives, on both branches,
\begin{equation}
\label{eq:mdww-laurent-j1}
a_1=\frac{H}{q},\qquad b_1=0.
\end{equation}

\paragraph{Resonance \(j=2\).}
The first row of \(M_2\) vanishes. The corresponding residual
coefficient is
\begin{equation}
\label{eq:mdww-laurent-obstruction}
[z^{-2}]L_1
=-Pa_{0,T}-a_0P_T+qa_0a_{1,y}
=-Pa_{0,T},
\end{equation}
where \(a_{1,y}=H_y/q=P_T/q\). The obstruction is therefore
\(2Pq'/q^2\) on the double-pole branch and \(-2Pq'/q^2\) on
the regular branch. Since \(P\ne0\), compatibility requires
\(q'\equiv0\) on the coefficient germ. A nonconstant germ
centred at an isolated zero of \(q'\) still has a nonzero
obstruction as a germ.

For the remaining resonances, take \(q\) constant. The second
residual at \(j=2\) determines \(b_2\) and leaves \(a_2\)
arbitrary. Specifically,
\begin{equation}
\label{eq:mdww-laurent-j2}
b_2=
\begin{cases}
Pa_2+a_{1,y},&\text{double pole in } V,\\
0,&\text{regular } V.
\end{cases}
\end{equation}

\paragraph{Double resonance \(j=3\).}
At order \(z^{-1}\), every term with an outer \(x\)-derivative
has zero Laurent coefficient, since \(D_x=\partial_z\).
The remaining term \(D_T(U_y)\) in the first residual also
has zero coefficient at this order, since \(a_{0,y}=0\) and
\(U_y\) has no \(z^{-1}\) term. The second residual reduces
to \(b_{1,T}=0\), which follows from
\eqref{eq:mdww-laurent-j1}. Hence \(a_3\) and \(b_3\) can
be prescribed independently.

\paragraph{Resonance \(j=4\).}
On the double-pole branch, write \(B=Pa_2+a_{1,y}=b_2\).
Both residual equations reduce to
\begin{equation}
\label{eq:mdww-laurent-j4-pole}
2Pa_4-4b_4+B_T-qa_2B=0.
\end{equation}
On the regular branch, the second residual reduces to
\((H-qa_1)b_3=0\), which follows from
\eqref{eq:mdww-laurent-j1}. The first residual gives one
linear relation between \(a_4\) and \(b_4\), with coefficients
\(10P\) and \(-4\), respectively. Thus, on either branch,
one may prescribe \(a_4\) arbitrarily and solve for \(b_4\).

For \(j\ge5\), \(M_j\) is invertible. Each compatible branch
therefore admits five freely prescribable analytic functions
of \((T,y)\),
\[
\psi,\qquad a_2,\qquad a_3,\qquad b_3,\qquad a_4,
\]
subject to \(\psi_y\ne0\). Once these functions are prescribed,
the recurrence determines all remaining coefficients.

\subsection{Preservation under admissible point transformations}

Equivalence transformations provide a way to compare singularity
tests across different coefficient choices
\cite{vaneeva2014equivalence}. Here the preservation of the formal
Laurent branches follows directly from the admissible maps in
\eqref{eq:mdww-general-point-map}.

Under a real-analytic admissible transformation, the image of
\(z=0\) has the Kruskal defining function
\[
\widetilde z=\widetilde x+\widetilde\psi(\widetilde T,\widetilde y),
\qquad
\widetilde\psi\bigl(F(T),Y(y)\bigr)=s(T)\psi(T,y)-h(T).
\]
Consequently,
\[
\widetilde z=s(T)z,\qquad
\widetilde\psi_{\widetilde y}=\frac{sP}{Y'}\ne0.
\]
The field transformation multiplies \(U\) and \(V\) by
nowhere-zero analytic factors and adds an analytic term to
\(\widetilde U\). Thus it preserves the pole order in each
field and maps each Laurent branch to the corresponding
branch of the target system. The inverse map has the same
properties, so the formal Laurent data transfer in both
directions. Moreover, \eqref{eq:mdww-coefficient-action} gives
\[
\widetilde q_{\widetilde T}(F(T))
=\frac{q'(T)}{kF'(T)}.
\]
Together with the nonvanishing of \(q\), \(P\), \(k\), and
\(F'\), this shows that the vanishing of
\eqref{eq:mdww-laurent-obstruction} is invariant.

\begin{theorem}
\label{thm:mdww-laurent-compatibility}
For the normalized MDWW class \eqref{eq:mdww-normalized}, both
integer-pole branches on movable analytic hypersurfaces with
\(\phi_x\phi_y\ne0\) are formally Laurent compatible if and
only if \(q\) is constant as a germ. In the original variables,
this is equivalent to \(\alpha/\beta=\mathrm{const}\).
\end{theorem}

The coefficient condition agrees with that obtained by Liu
and Gao \cite[Eqs.~(9)--(10)]{liu2023painleve}. The calculation
above identifies the obstruction at \(j=2\) on both branches
and establishes the freedom of the coefficients at every
positive resonance. The conclusion is restricted to formal
compatibility on the stated hypersurfaces. Convergence and
the analysis of hypersurfaces with \(\phi_x\phi_y=0\) are
separate questions in nonlinear singularity analysis
\cite{kruskal1997analytic}.

By Corollary~\ref{cor:mdww-standard-equivalence}, the formal
Laurent-compatible locus coincides with the locus of local
point equivalence to standard MDWW. This common locus is
strictly contained in the locus admitting additional time
symmetry. For instance, \(q=1+T\) near \(T=0\) admits
\[
(T+1)\partial_T+\tfrac12x\partial_x
-\tfrac32(U\partial_U+V\partial_V)
\]
by \eqref{eq:mdww-lie-regular-canonical}. Its resonance
obstruction is \(2P/(1+T)^2\) on the double-pole branch and
\(-2P/(1+T)^2\) on the regular branch.
\section{Comparison of point equivalence, time symmetries, and Laurent compatibility}
\label{sec:mdww-loci}

Equivalence transformations provide a framework for comparing
reducibility to constant coefficients with conditions obtained
from singularity analysis \cite{vaneeva2014equivalence}. For the
normalized MDWW class, we compare these two conditions with
the Lie symmetry extensions classified in
Section~\ref{sec:mdww-lie-classification}.

Let \(\mathscr A\) denote the set of pointed real-analytic
coefficient germs \([q]_{T_0}\) with \(q(T_0)\ne0\). The
equivalence and symmetry statements below are local at the
base time \(T_0\). Define
\begin{equation}
\label{eq:mdww-loci}
\begin{aligned}
\mathcal P
&=\bigl\{[q]_{T_0}\in\mathscr A:
  \mathcal E_q \text{ is locally point-equivalent to }\mathcal E_2\bigr\},\\
\mathcal S
&=\bigl\{[q]_{T_0}\in\mathscr A:
  \mathfrak g_q\supsetneq\mathfrak g^\cap\bigr\},\\
\mathcal L
&=\bigl\{[q]_{T_0}\in\mathscr A:
  \text{both Laurent branches are formally compatible}\bigr\}.
\end{aligned}
\end{equation}
Here \(\mathfrak g^\cap\) is the kernel algebra
\eqref{eq:mdww-lie-kernel}, obtained by intersecting the maximal
Lie invariance algebras in the fixed coordinates \((T,x,y,U,V)\).
The Laurent branches are those of Section~\ref{sec:mdww-laurent}
on movable analytic hypersurfaces with \(\phi_x\phi_y\ne0\).
They are distinguished by whether \(V\) has a double pole or
is regular.

Every member of the class admits the family \(X_q(\lambda)\)
in \eqref{eq:mdww-lie-baseline}. A choice with
\(\lambda'\not\equiv0\) gives a generator outside
\(\mathfrak g^\cap\), so \(\mathcal S=\mathscr A\). To
distinguish the germs that admit a symmetry with a nonzero
time component, define
\begin{equation}
\label{eq:mdww-time-symmetry-locus}
\mathcal S_{\mathrm{time}}
=\bigl\{[q]_{T_0}\in\mathscr A:
  \mathfrak g_q\text{ contains a generator with }
  \tau\not\equiv0\bigr\}.
\end{equation}
Equivalently, \([q]_{T_0}\in\mathcal S_{\mathrm{time}}\) if and
only if \(\mathfrak g_q\supsetneq\mathfrak b_q\), where
\(\mathfrak b_q\) in \eqref{eq:mdww-lie-baseline} consists
of the generators with zero time component.

\begin{proposition}
\label{prop:mdww-loci}
Let \(\mathscr C\) be the set of constant coefficient germs
in \(\mathscr A\), and let
\[
\mathscr R=\bigl\{[q]_{T_0}\in\mathscr A:q'(T_0)\ne0\bigr\}
\]
be the set of regular coefficient germs. Then
\begin{equation}
\label{eq:mdww-loci-classified}
\mathcal P=\mathcal L=\mathscr C,\qquad
\mathcal S_{\mathrm{time}}=\mathscr C\cup\mathscr R,\qquad
\mathcal S=\mathscr A.
\end{equation}
In particular,
\[
\mathcal P=\mathcal L
\subsetneq\mathcal S_{\mathrm{time}}
\subsetneq\mathcal S.
\]
\end{proposition}

\begin{proof}
Corollary~\ref{cor:mdww-standard-equivalence} gives
\(\mathcal P=\mathscr C\). Let \(\mathcal L_{\mathrm{pole}}\)
and \(\mathcal L_{\mathrm{reg}}\) denote the compatibility
loci for the double-pole and regular branches of \(V\),
respectively. The \(j=2\) obstructions in
\eqref{eq:mdww-laurent-obstruction} are
\(2\psi_y q'/q^2\) and \(-2\psi_y q'/q^2\), respectively.
Since \(\psi_y\ne0\) and \(q\ne0\), either obstruction
vanishes as a germ if and only if \(q'\equiv0\). For constant
\(q\), all remaining resonance conditions are compatible on
both branches by Theorem~\ref{thm:mdww-laurent-compatibility}.
Consequently,
\[
\mathcal L_{\mathrm{pole}}
=\mathcal L_{\mathrm{reg}}
=\mathcal L
=\mathscr C.
\]
This common compatibility locus is invariant under the
real-analytic admissible point transformations, which
preserve both branches and the hypersurface conditions in
Section~\ref{sec:mdww-laurent}.

Theorem~\ref{thm:mdww-maximal-lie-algebras} gives
\(\mathcal S_{\mathrm{time}}=\mathscr C\cup\mathscr R\),
and the family \(X_q(\lambda)\) gives
\(\mathcal S=\mathscr A\). At \(T_0=0\), the germ \(q=1+T\)
belongs to \(\mathcal S_{\mathrm{time}}\setminus\mathcal L\),
and the germ \(q=1+T^2\) belongs to
\(\mathcal S\setminus\mathcal S_{\mathrm{time}}\).
These two germs establish the strict inclusions.
\end{proof}

The equality \(\mathcal P=\mathcal L\) identifies the
constant-ratio Laurent condition
\(\alpha/\beta=\mathrm{const}\) obtained by Liu and Gao
\cite[Eqs.~(9)--(10)]{liu2023painleve} with the condition for
local point equivalence to standard MDWW. The regular germ
\(q=1+T\) at \(T_0=0\) illustrates how these conditions differ
from the existence of a time symmetry:
\[
\begin{array}{c|c}
\text{Additional time generator}
&
\text{Obstruction at }j=2
\\[1mm] \hline
\displaystyle
(T+1)\partial_T+\tfrac12x\partial_x
-\tfrac32(U\partial_U+V\partial_V)
&
\displaystyle
\begin{cases}
 2\psi_y/(1+T)^2,&\text{double pole in }V,\\[1mm]
-2\psi_y/(1+T)^2,&\text{regular }V.
\end{cases}
\end{array}
\]
Thus the time generator coexists with a nonzero Laurent
obstruction on both branches. For the critical germ
\(q=1+T^2\) at \(T_0=0\), the obstruction vanishes at the
base time alone. Since \(q'=2T\), it is nonzero as a germ,
and both branches fail the formal compatibility condition.
\section{Exact solutions on the equivalence and symmetry branches}
\label{sec:mdww-exact-solutions}

Exact solutions of MDWW systems have been constructed by several
methods, including consistent Riccati expansions
\cite{liang2019consistent}, rational-solution constructions
\cite{ren2019rational}, and bilinear methods for the time-dependent
system \cite{liu2023painleve}. Here we use three explicit solutions
to illustrate the coefficient classification. The first is a
pullback from standard MDWW, the second belongs to a regular
nonconstant coefficient germ, and the third exhibits a movable
pole on a hypersurface with \(\phi_y=0\).

\subsection{Pullback from standard MDWW}

Let \(z=x+y+T\). The fields
\begin{equation}
\label{eq:mdww-standard-pole-solution}
\widehat U=\frac{1}{z}+\frac12,\qquad
\widehat V=-\frac{1}{z^2}
\end{equation}
solve the standard system \(\mathcal E_2\) wherever \(z\ne0\).
Indeed, with \(z=x+y+T\), both time-derivative terms are
\(2z^{-3}\), the dispersive terms in each residual sum to
\(6z^{-4}\), and the nonlinear terms contribute
\(-2z^{-3}-6z^{-4}\).

For the original system \eqref{eq:mdww-original}, take
\[
\alpha(t)=\beta(t)=e^t,\qquad T=e^t-1.
\]
The time normalization \eqref{eq:mdww-time-normalization} with
\(t_0=0\) gives \(q=1\). The field rescaling
\(U=2\widehat U\), \(V=2\widehat V\) carries
\eqref{eq:mdww-standard-pole-solution} to a solution of
\(\mathcal E_1\). Returning to the original time yields
\begin{equation}
\label{eq:mdww-pullback-solution}
u(t,x,y)=1+\frac{2}{x+y+e^t-1},\qquad
v(t,x,y)=-\frac{2}{(x+y+e^t-1)^2}.
\end{equation}
Each residual of \eqref{eq:mdww-original} is \(2e^t\) times the
corresponding residual of \(\mathcal E_2\) evaluated at
\(T=e^t-1\). Hence \eqref{eq:mdww-pullback-solution} solves the
original system wherever \(x+y+e^t-1\ne0\). Both original
coefficients vary with time, and their constant ratio places
the system in the standard-equivalence locus of
Corollary~\ref{cor:mdww-standard-equivalence}.

\subsection{An exact solution on the regular nonconstant branch}

Consider the coefficient \(q(T)=1+T\) near \(T=0\). This germ
admits the additional time generator
\eqref{eq:mdww-lie-regular-canonical} and is point-inequivalent
to standard MDWW by
Corollary~\ref{cor:mdww-standard-equivalence}.
To construct a solution with both nonlinear terms active,
use the ansatz
\[
U=A(T)(x+y),\qquad V=C(T)(x+y).
\]
The derivatives and nonlinear terms in
\eqref{eq:mdww-normalized} are
\[
\begin{aligned}
U_{Ty}&=A',&
U_{xxy}&=V_{xx}=0,&
(UU_y)_x&=A^2,\\
V_T&=C'(x+y),&&&
(UV)_x&=2AC(x+y).
\end{aligned}
\]
The residuals therefore reduce to
\[
L_1=A'-qA^2,\qquad
L_2=(C'-2qAC)(x+y).
\]
For \(q=1+T\), the choice \(A=-2q^{-2}\) and \(C=q^{-4}\)
gives
\[
A'=4q^{-3}=qA^2,\qquad
C'=-4q^{-5}=2qAC.
\]
It follows that
\begin{equation}
\label{eq:mdww-nonconstant-polynomial-solution}
U=-\frac{2(x+y)}{(1+T)^2},\qquad
V=\frac{x+y}{(1+T)^4}
\end{equation}
solves both equations for \(T>-1\) and \(x,y\in\mathbb R\).
On this domain, \((UU_y)_x\) is nowhere zero and \((UV)_x\)
is nonzero wherever \(x+y\ne0\). The solution is analytic
near \(T=0\). Its denominators vanish at \(T=-1\), where
the coefficient \(q\) also vanishes.

\subsection{Pole solutions on \texorpdfstring{\(y\)}{y}-independent hypersurfaces}
\label{subsec:mdww-exceptional-poles}

Let \(q(T)\) be analytic and nowhere zero on a time interval
\(I\). Choose analytic functions \(a(T)\), \(B(T)\), \(\psi(T)\),
and \(g(y)\), and set
\[
z=x+\psi(T),\qquad
E(T)=\exp\!\left(2\int_{T_*}^{T}q(s)B(s)\,\mathrm{d}s\right),
\qquad T_*\in I.
\]
Then
\begin{equation}
\label{eq:mdww-exceptional-pole-solution}
U=\frac{a(T)}{z}+\frac{\psi'(T)}{q(T)}+B(T)z,
\qquad
V=g(y)E(T)z
\end{equation}
solves \eqref{eq:mdww-normalized} wherever \(z\ne0\).
Since \(U_y=0\) and \(V_{xx}=0\), the first equation is
satisfied identically. For the second equation, the product
\[
UV=g(y)E(T)\left(a(T)+\frac{\psi'(T)}{q(T)}z+B(T)z^2\right)
\]
gives
\[
q(T)(UV)_x
=g(y)E(T)\bigl(\psi'(T)+2q(T)B(T)z\bigr)
=V_T,
\]
where \(E'=2qBE\) and \(z_T=\psi'\).

Where \(a(T)\ne0\), the field \(U\) has a simple pole on the
movable hypersurface \(x=-\psi(T)\), and \(V\) extends
analytically across it. The defining function \(z\) satisfies
\(z_x=1\) and \(z_y=0\), so this family lies outside the
hypersurface condition \(\phi_x\phi_y\ne0\) in
Theorem~\ref{thm:mdww-laurent-compatibility}. The residue
\(a(T)\) is arbitrary in \eqref{eq:mdww-exceptional-pole-solution},
whereas the regular-\(V\) balance under \(\phi_x\phi_y\ne0\)
fixes the residue at \(-2/q(T)\). This construction gives
\(y\)-independent movable pole hypersurfaces for arbitrary
admissible \(q\), including nonconstant germs. The general
Laurent recurrence and the count of free coefficient
functions on these hypersurfaces remain to be determined.
% --------------------------------------------------
% Appendices (uncomment if required)
% --------------------------------------------------
%\begin{appendices}
%\section{Supplementary calculations}
%\label{app:calculations}
% Insert supplementary calculations here.
%\end{appendices}

% --------------------------------------------------
% Declarations
% --------------------------------------------------
\section{Conclusion}

The equivalence groupoid of the normalized MDWW class acts on its
coefficient by positive time reparametrization and nonzero constant
rescaling. The coefficient action is realized by maps induced from
the broader coupled class, and the direct MDWW proof shows that it
exhausts all local point identifications. For nonconstant pointed
analytic germs, the order of the first nonzero derivative and its
sign relative to the base coefficient value form a complete pair
of invariants.

Local point equivalence to standard MDWW and formal Laurent
compatibility on both tested branches select exactly the constant
coefficient germs. Regular nonconstant germs additionally admit
one independent time generator, whereas nonconstant germs centred
at a critical time admit no such generator through that time.
The exact pole family with \(\phi_y=0\) illustrates the role of the
hypersurface hypothesis. The general Laurent recurrence on those
hypersurfaces and convergence of the formal series remain open
in this analysis.

\section*{Declarations}

\subsection*{Author contributions}
M.S. is the sole author and was responsible for the
conceptualization, methodology, formal analysis, investigation,
validation, and writing, review, and editing of the manuscript.

\subsection*{Funding}
The author received no financial support for this research.

\subsection*{Data availability}
No datasets were generated or analysed during the current study.

\subsection*{Conflict of interest}
The author declares no conflict of interest.

\subsection*{Declaration of generative AI use}
During the preparation of this manuscript, the author used
ChatGPT (OpenAI) to assist with LaTeX formatting, language editing,
and the drafting and symbolic checking of mathematical derivations.
The author takes full responsibility for validating all mathematical
statements and for the final content of the manuscript.
% --------------------------------------------------
% References
% --------------------------------------------------
% Add references.bib and uncomment the following lines when needed.
% \bibliography{references}
% \bibliographystyle{plain}

\end{document}